# Bipolar-driven large magnetoresistance in silicon

Michael P. Delmo*, Eiji Shikoh, Teruya Shinjo, and Masashi Shiraishi

*Graduate School of Engineering Science, Osaka University*

*1-3 Machikaneyama, Toyonaka, Osaka, 560-8531, Japan*

Large linear magnetoresistance (MR) in electron-injected p-type silicon at very low magnetic field is observed experimentally at room temperature. The large linear MR is induced in electron-dominated space-charge transport regime, where the magnetic field modulation of electron-to-hole density ratio controls the MR, as indicated by the magnetic field dependence of Hall coefficient in the silicon device. Contrary to the space-charge-induced MR effect in unipolar silicon device, where the large linear MR is inhomogeneity-induced, our results provide a different insight into the mechanism of large linear MR in non-magnetic semiconductors that is not based on the inhomogeneity model. This approach enables homogeneous semiconductors to exhibit large linear MR at low magnetic fields that until now has only been appearing in semiconductors with strong inhomogeneities.

The presence of inhomogeneities [1-3], such as, macroscopic inclusions [4], defects [5-8], and electric field fluctuations [9-14] in non-magnetic materials generates magnetoresistance (MR) effect - the relative change of resistance (*R*) due to application of magnetic field (*H*) - that does not saturate with increasing *H* [1-3]. If the inhomogeneity is strong, this leads to a large MR that shows linear *H*-dependence, as demonstrated by Parish and Littlewood in their classical model of inhomogeneous conductor [2,3]. Large linear MR is best exhibited by doped silver chalcogenides ($Ag_{2+\delta}Se$ and $Ag_{2+\delta}Te$), where the MR is linear from ~10 mT up to 55 T, without showing any sign of saturation even at room temperature [5]. Recently, Delmo *et al.* [9,10] demonstrated that a simple two-terminal silicon device exhibits large linear MR, when the transport is space-charge-limited [9-13], where the space-charges induce spatially fluctuating electric field (*E*), which generates the inhomogeneity [9]. Here, they show a different kind of disorder, one that is not inherent to the material, but an inhomogeneous *E* that can be introduced and tuned externally by bias voltage [9,10]. This charge injection approach has been adopted by Wan *et al.* [14] to geometrically enhance the MR of four-terminal silicon device by injecting holes into n-type silicon in low magnetic fields. Their experimental results support the inhomogeneity model [2,3], suggesting that the boundary between the hole- and electron-dominant conduction regions - the p-n boundary - provides the inhomogeneity that induces large linear MR at room temperature [14].

In general, large linear MR in non-magnetic materials, especially the ones showing at room temperature, is strongly associated with inhomogeneity. Although, quantum routes to large linear MR has been proposed by Abrikosov [15], his model is restricted to semimetals [16,17] and narrow-gap semiconductors [18], including graphene [19]. So, is inhomogeneity the only classical route that leads to large linear MR, particularly in low magnetic fields? In this Letter, we demonstrate that when electrons, in the form of space-charges, are injected into p-type silicon device large linear MR is induced at low $H$, at 300 K. We show, via simple classical model and simulation, that the magnetic field modulation of the electron-to-hole density ratio (EHR) is the origin of the large linear MR, as indicated by the $H$-dependence of Hall coefficient ($R^H$) of the silicon device. This approach enables homogeneous semiconductors to exhibit large linear MR at low $H$, even without the presence of strong inhomogeneities.

In the experiment, we used ultralow boron-doped, p-type silicon [p-Si (001)] substrates (Nilaco Corporation) with thickness, $t = 0.5$ mm and resistivity, $\rho = 90$ $\Omega$-m, which has carrier density, $n_h = 2.0 \times 10^{12}$ cm$^{-3}$ and hole mobility, $\mu_h = 350$ cm$^2$V$^{-1}$s$^{-1}$ at 300 K, measured by van der Pauw method [10]. We followed the methods used in Ref. [9] to fabricate our p-Si devices, where the metal electrode is Indium, In, and the In/p-Si contact is ohmic, as shown in Fig. 1(a). We measured the two-terminal current ($I$)-voltage ($V$) characteristics, Hall voltages ($V_H$), and four-terminal voltages ($V_{FT}$) of the device in a current-in-plane geometry with $H$ applied perpendicular to the substrate plane.

The MR of the p-Si device at low $H$ can be enhanced effectively by increasing the bias voltage ($V_{Bias}$). Figure 1(b) shows the MR as a function of $H$ from -5 to 5 T, at given $V_B$ at 300 K. The magnetoresistance ratio is defined as MR $= \Delta R/R = [R(H) - R(H = 0)] / R(H = 0) \times 100$ %, where $R = V/I$. The MR below $|H| = 1$ T shows considerable enhancement when the $V_{Bias}$ is increased from 1 to 200 V. For low $V_{Bias}$ (1 ~ 10 V), the MR is typically small and shows $H^2$-dependence [23,24]. For example, MR ≈ 0.13 % at $H$ = 250 mT and $V_{Bias}$ = 5 V. However, for high $V_{Bias}$ (200 V, red line), MR is significantly enhanced, resulting in MR ≈ 15 % at 50 mT, 6.5 % at 25 mT, and 3 % at 15 mT [see also Fig. 1(c)]. Figure 1(c) shows MR in $H$ = 0 ~ 250 mT (dotted red line) at 200 V, which clearly shows MR enhancement at low $H$. The linear fit (blue line) indicates that the MR exhibits linear response to $H$ that extends down to ~ 5 mT.

To explore the origin of the large linear MR effect at low $H$, we measured the $I$-$V$ characteristics of the device for various $H$ at 300 K, as shown in Fig. 2(a) in $\log_{10}$-$\log_{10}$ scale. For $H$ = 0 T, the $I$-$V$ curve shows linear behavior below $V_{Bias}$ = 10 V, indicating ohmic transport, whereas at $V_B$ = 40 ~ 80 V the $I$-$V$ shows an intermediate region, where $I \propto V^{1/2}$. As $V_{Bias}$ increases further, particularly above 150 V, $I \propto V^2$ is observed. This kind of transport behavior is explained as characteristic of electron-injected p-type semiconductor, where holes control $I$ at ohmic regime, but transforms into electron-dominated $I$ at high $V_{Bias}$ regime [25]. We verified this charge carrier reversal by measuring $R^H$ as a function of $V_{Bias}$, as shown in Fig. 2(b) [dotted red line]. The sign of $R^H$ is positive at low $V_{Bias}$ regime but reverses to negative above $V_{Bias}$ ≈ 10 V, which clearly shows that the dominant charge carrier changes from holes to electrons. This result indicates that electrons are effectively injected into the device at high $V_{Bias}$, at 0 T [23,26]. Furthermore, the distinctive $I$-$V$ characteristic, particularly the $I \propto V^2$ regime, is caused by space-charge effect, as explained in Ref. [25]. To verify this, we estimated the Debye length, $\lambda_D = (\varepsilon_{Si}\varepsilon_0 k_B T/q^2 n_{excess})^{1/2}$ = 6.62 μm of the p-Si device at 150 V and 300 K, where $\varepsilon_{Si}$ = 12.0 is the relative permittivity of silicon [26], $\varepsilon_0$ is the vacuum permittivity, $k_B$ is the Boltzmann constant, $T$ is the temperature, $q$ is the electron charge, and $n_{excess}$ is the excess electron density. We calculated $n_{excess} = 3.5 \times 10^{11}$ cm$^{-3}$ from $R^H$ in Fig. 2(b), using $R^H = [q(n_e - n_h)]^{-1}$, where $n_e$ is the electron density and $n_e - n_h = n_{excess}$, because holes also contribute significantly to the transport [23]. We also calculated the average distance, $d = 1/(n_{excess})^{1/3} = 1.37$ μm between the excess electrons. Since, $\lambda_D > d$, this indicates that electrons are correlated via unscreened Coulomb interaction within the Debye length [26]. Thus, the quasi-neutrality is broken in the device, which indicates space-charge effect at zero magnetic field [9-11,26]. We also verified the space-charge effect experimentally by measuring the spatial dependence of carrier density in the device [see Fig. 3 later].

As $H$ increases from 0 T, the $I$-$V$ shows monotonously decreasing $I$ at a fixed $V_{Bias}$, resulting in positive MR, as shown in Fig. 2(a). However, the $I$-$V$ is strongly affected by $H$ in non-ohmic regime ($V_{Bias} > 10$ V), particularly at $I \propto V^2$ regime ($V_B > 150$ V), than in ohmic regime ($V_B < 10$ V). To see this difference clearly, we plot the MR ($H = 0.5$ T), as a function of $V_{Bias}$, as shown in Fig. 2(b) (dotted blue line). $H = 0.5$ T was chosen to tract the effect of the carrier-type reversal (dotted red line) on the MR. In hole-dominated ohmic regime ($V_{Bias} = 1 \sim 10$ V), the MR is small and shows weak $V_{Bias}$-dependence (MR ≈ 0.3 % at 10 V), whereas the MR rapidly increases, as $V_{Bias}$ increases from 10 V up to 40 V (MR ≈ 2.6 % at 40 V). In the $I \propto V^{1/2}$ regime (40 ~ 80 V), however, MR again shows relatively weak $V_{Bias}$-dependence (MR ≈ 3.6 % at 80 V). Surprisingly, MR is enhanced significantly in the electron-dominated $I \propto V^2$ regime, where MR ≈ 140 % at 200 V.

The induction of large linear MR at high $V_{Bias}$ in the device at low $H$ can be attributed to space-charge effect [9,10], assuming that only excess electrons, in the form of space-charges, induce the MR [9]. In this case, the inhomogeneity model can be used to estimate the important features of space-charge-induced MR effect [2,3,9]. At 200 V, MR $\propto \mu_e H \approx 2.6$ % at 250 mT for $\mu_e \approx 3\mu_h = 1,050$ cm$^2$V$^{-1}$s$^{-1}$, which is about thirty times smaller than the experimental value (MR $\approx 77$ %), where $\mu_e$ is electron mobility. Furthermore, the quadratic-to-linear MR crossover field, $|H_c| \approx \mu_e^{-1} = 9.5$ T, is far larger than the experimental value of $|H_c| \leq 5$ mT [Fig. 1(c)]. These results suggest that the MR is likely controlled by strong inhomogeneity that is characterized by the distribution width of mobility, $\Delta\mu$, the measure of mobility disorder [2,3]. However, $\Delta\mu \approx H_c^{-1} = 2 \times 10^6$ cm$^2$V$^{-1}$s$^{-1}$ is extremely large to be realistic for silicon at room temperature, because in unipolar space-charge effect, the MR is characterized by the average mobility, $<\mu>$, as indicated in Ref. [9] and [10]. These results suggest that the contribution of space-charge-induced inhomogeneity to the large linear MR is negligibly small. We note that the rapid increase of MR at low $H$ cannot be associated to avalanche breakdown, as reported by Sun *et al*. [27] and Schoonus *et al*. [28] because at $V_{Bias} = 200$ V, $E \approx 330$ Vcm$^{-1}$ is three orders of magnitude lower than the breakdown field of silicon [29].

In contrast, the results in Fig. 2(a) and 2(b) indicate that the transport in high $V_{Bias}$ is bipolar. Only few mechanisms are known to induce large positive MR in bipolar-injected semiconductor device at low $H$ [14,30,31]. For example, the p-n-boundary [5] in hole-injected n-type silicon can induce large linear MR, but in this case it is important that the MR is measured via four-terminal method. We verified via four-terminal method that the p-n boundary is not the mechanism of the large linear MR in our silicon devices (see Supplemental Material, SM [30]). In addition, it is known that electron injection into p-type indium antimonide (InSb) exhibits large MR at low $H$, in which the deflection of electron-hole plasma generates the effect [31,32]. Here it is important that $\mu_e \gg \mu_h$, (for InSb, $\mu_e \approx 64\mu_h$ [29]), and Suhl effect - the deflection of electrons and holes on the same surface of the device - is generated [33]. Although, we verified existence of the plasma in our p-Si device by measuring negative resistance - decreasing $I$ with increasing $V$ at 300 K [20-22,31,32], the deflection of the plasma by $H$ is not the likely mechanism here, because $\mu_e \approx 3\mu_h$ in silicon and Suhl effect was not observed (SM [30]).

Figure 2(c) shows $R^H$ and two-terminal MR of the device as a function of $H$ at 200 V (measured simultaneously). Surprisingly, $R^H$ curve (dotted red curve) agrees perfectly with the MR curve (dotted blue curve). At $H = 0$ T, $R^H$ is negative, which indicates that electrons are the dominant charge carrier. As $H$ increases from 0 to 240 mT, $R^H$ increases dramatically from -71.5 to 0 $m^3C^{-1}$, but above 250 mT $R^H$ changes to positive (hole dominate the transport) and increases to 18.3 $m^3C^{-1}$ at 400 mT, but with a different $H$-response to that of the negative $R^H$. These results indicate that EHR can be modulated by $H$, and therefore suggest that EHR causes the large linear MR. It is known that MR can be enhanced by modulating EHR in bipolar semiconductors, for example, by changing the concentration of acceptor and donor impurities [34], or by application of pressure [35].

To support our conclusion that the EHR is $H$-modulated, we measured $R^H$ as a function of $H$ at three different positions of the Hall voltage probes (A, B, and C) in p-Si device (sample IP6) in the space-charge regime, as shown in Fig. 3(a). We verified the space-charge effect by measuring carrier density ($n$) as a function of probe position ($x$), as shown in Fig. 3(b). At 1 V, $n$ is almost independent of all values of $x$ (blue dots), which indicates Ohmic transport, whereas at 100 V, $n$ is strongly dependent on $x$ (red dots), where the fit (red line) shows that $n \propto x^{-1/2}$ in agreement with the Mott-Gurney theory, indicating space-charge effect [36]. We note that holes and electrons dominate the transport at 1 V and 100 V, respectively, similar to the result in Fig. 2(b) (SM [30]).

The colored curves in Fig. 3(c) - 3(e) show $R^H$ [left axis] as a function of $H$ [bottom axis] for A, B, and C, respectively from 0 to 5 T. The results show that $R^H$ is strongly dependent on $H$, as well as on $x$. The fact that $R^H$ is strongly modulated by $H$ indicates that $H$ changes EHR in the device. We calculated $R^H$ as a function of $n_e/n_h$ (EHR), which is expressed by $R^H = (1/qn_h) \times (1 - n_r\mu_r^2)/(1 + n_r\mu_r)^2$, where $n_r = n_e/n_h$ and $\mu_r = \mu_e/\mu_h = 3$, which is just the conventional formula for bipolar Hall effect [23,26,37] (see SM [30] for model and calculations). The results of the simulation are the fit curves in black in Fig. 3(c) - 3(e), where the abscissa is $(n_e/n_h)^{-1}$ [top axis] and the ordinate is normalized $R^H$ ($= qn_hR^H$) [right axis]. Surprisingly, the fits agree perfectly with the experimental results for all three Hall voltage probes, even the negative-to-positive $R^H$ reversal field is reproduced in Fig. 3(d) and 3(e). Thus, the simulation suggests that the $H$-modulated $R^H$ is simply the result of EHR modulation by $H$. Similarly, we fit the simulated $qn_hR^H$ vs. $(n_e/n_h)^{-1}$ curve (black line) to the $R^H$ vs. $H$ curves (dotted red line) of sample IP3, as shown in Fig. 4. Indeed, the fit reproduced the behavior of $R^H$ below 400 mT, which corroborates our conclusion that the magnetic field modulation of EHR generates the large linear MR (dotted blue line) in Fig. 2(c). But we note that the linear response of MR to $H$ is not reproduced in the simplified model, which suggests that the linearity is not an effect of EHR but of other factors, such as unscreened Coulomb interaction in space-charge effect [9], which we did not consider in the model. Furthermore, the dependence of $R^H$ vs. $H$ curve on $x$ suggests that the extent by which electrons propagate in the device is suppressed by increasing $H$, which implies that the modulation of EHR can be associated to the effect of $H$ to decrease the electron diffusion length either by lowering the electron mobility [31,32], or by reducing the electron lifetime [37].

In conclusion, we have demonstrated experimentally that by injecting electrons, in the form of space-charges, into p-type silicon device, large linear MR can be induced at low $H$. Our measurement and simulation suggest that the modulation of EHR by $H$ is the origin of the large linear MR. Although our results will help understand the mechanism of the large linear MR in homogeneous semiconductor device, the microscopic origin is still not clear, thus further studies will be necessary. Finally, the large linear MR, which can be tuned effectively by bias voltage and low magnetic fields, is also of considerable technological importance. The relative sensitivity, $S = (\Delta R/R)/H \approx 3.15$ T$^{-1}$ for $V_{Bias} = 200$ V and $|H| < 250$ mT [Fig. 1(c)], of the present silicon device is larger than those of commercially known semiconductor magnetic field sensors ($S = 0.07 \sim 3.0$ T$^{-1}$ for $H = 190$ mT) [38], which makes silicon a technologically attractive material for ultralow magnetic field sensing applications [38,39].

This work is supported by KAKENHI. We appreciate discussions with K. Kobayashi and I. Appelbaum. We thank Y. Suzuki for allowing us to use PPMS in his laboratory. M. P. D. acknowledges support from JSPS Postdoctoral Fellowship for Foreign Researchers.

*delmo@semi.ee.es.osaka-u.ac.jp

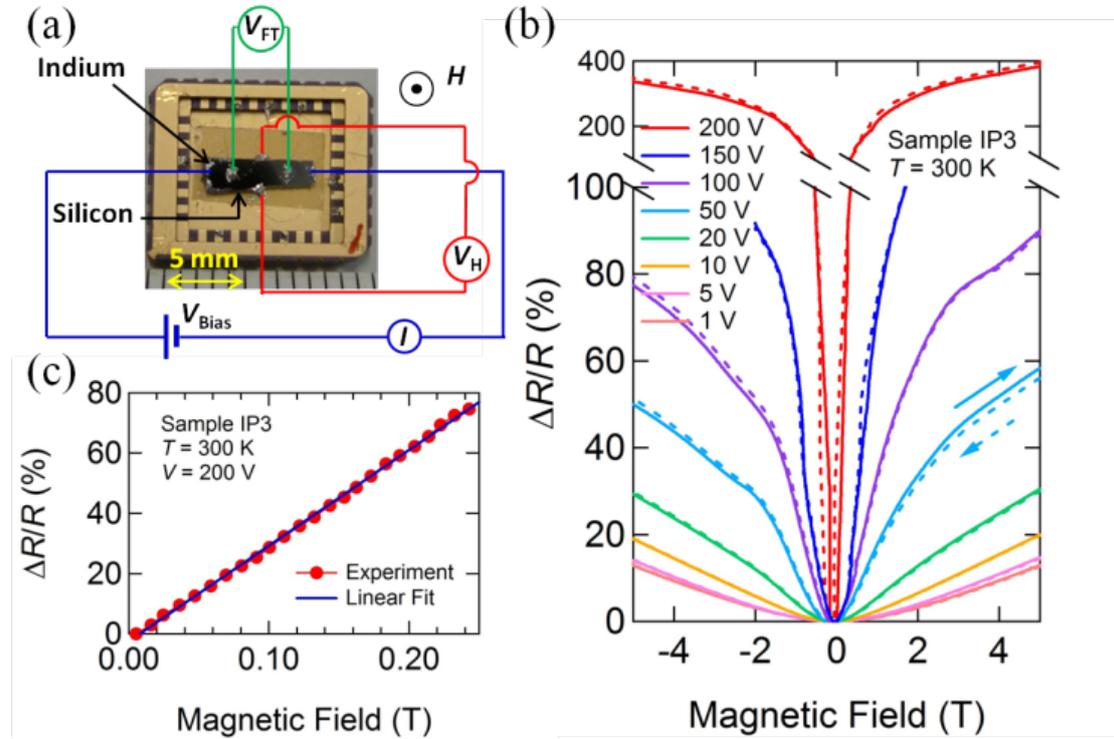

FIG. 1. Two-terminal MR of p-Si device at 300 K. (a) Photo of a typical p-Si device and the measurement schematic. The black rectangular specimen is the p-Si, the In electrodes connected by the blue circuit are the current injection-drain electrodes (for two-terminal MR measurements), the red circuit is for the Hall measurements, and the green circuit is for four-terminal MR measurements. The device dimension is $L = 6$ mm and $W = 2$ mm, where $L$ is the distance between current injection and drain electrodes and $W$ is the width. (b) MR as a function of $H$ from -5 to 5 T, for various $V_B$ from 1 to 200 V. The separated upper portion shows the MR at 200 V, where the scale of $\Delta R/R$ (100 ~ 400 %) is different to that of the lower portion. The arrows show the sweeping direction of $H$. (c) MR as a function of $H$ from 0 to 250 mT at $V_B = 200$ V (dotted red line). The linear fit (blue line) shows that the MR is linear, even in low $H$.

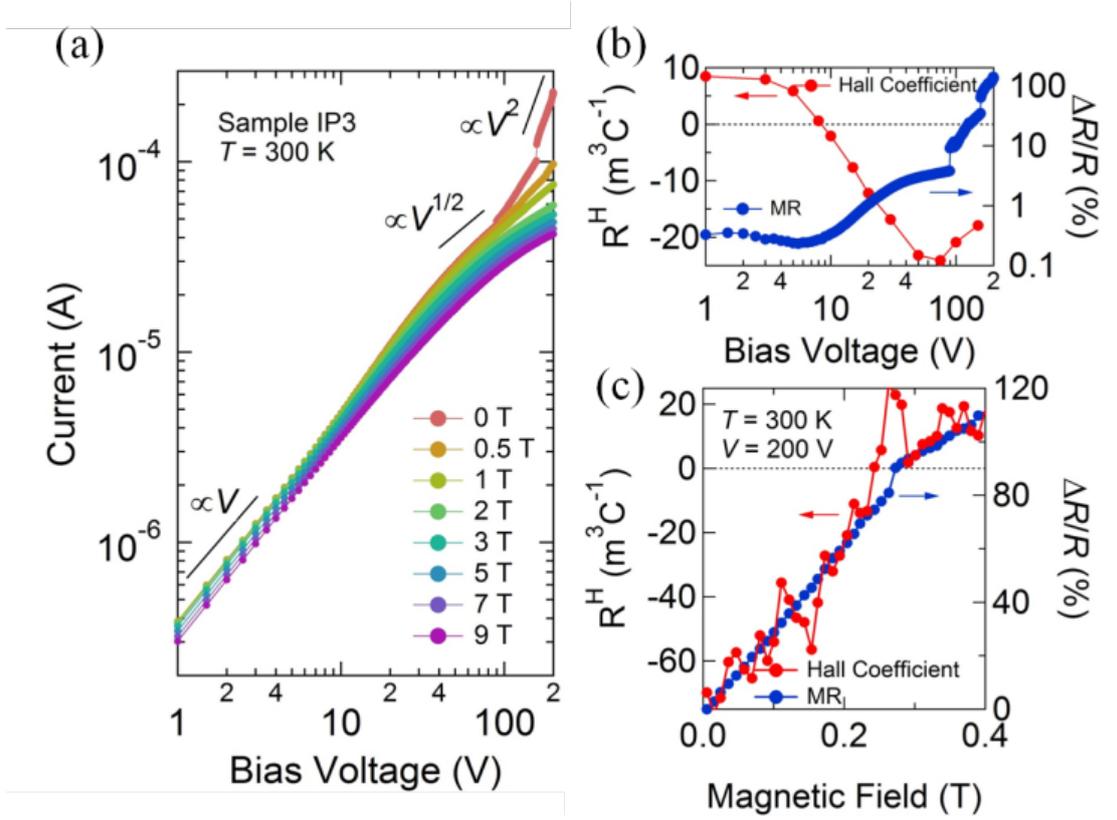

FIG. 2. *I-V* and Hall measurements of p-Si device at 300 K. (a) *I-V* characteristics of the p-Si device for various *H* from 0 to 9 T, plotted in $\log_{10}$-$\log_{10}$ scale. The three black lines above the *I-V* curves show the slope of the ohmic ($I \propto V$), intermediate regime ($I \propto V^{1/2}$), and space-charge ($I \propto V^2$) behavior. (b) $R^H$ as a function of $V_{Bias}$ (dotted red line) and two-terminal MR of the p-Si device as a function of $V_{Bias}$ (dotted blue line). (c) $R^H$ as a function of *H* (dotted red line) and two-terminal MR as a function of *H* (dotted blue line).

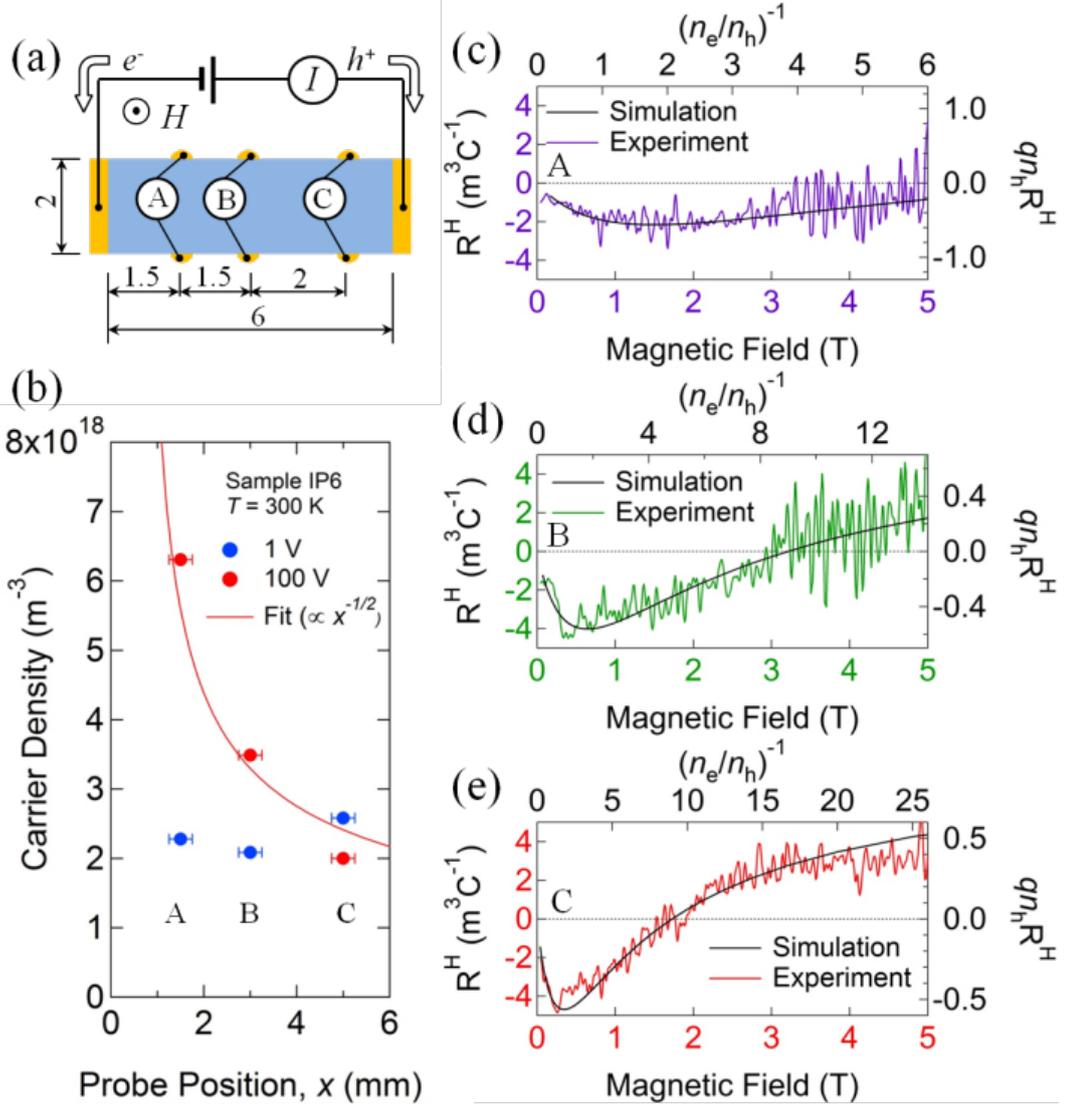

FIG. 3. Multi-probe measurement of $R^H$. (a) Schematic of p-Si device (sample IP6) with multiple Hall voltage probes (A, B, and C). Dimensions are in millimeter. (b) Carrier density as a function of probe position. Error bar is the width of the voltage probes (~0.5 mm). (c) - (e) $R^H$ as a function of $H$ at different probe positions (colored curves). Normalized $R^H$ as a function of $(n_e/n_h)^{-1}$ (black line).

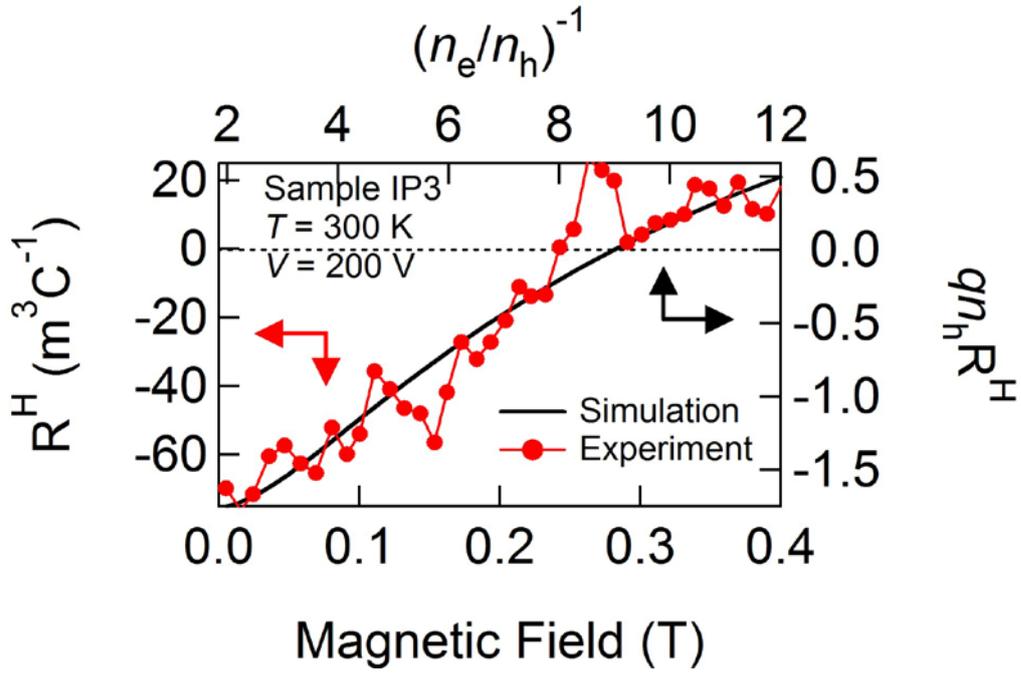

FIG. 4. $R^H$ as a function of $H$ (dotted red line), and normalized $R^H$ as a function of $(n_e/n_h)^{-1}$ (black line) of sample IP3.

*Supplemental Material*

**Bipolar-driven magnetoresistance in silicon**

Michael P. Delmo*, Eiji Shikoh, Teruya Shinjo, and Masashi Shiraishi

*Graduate School of Engineering Science, Osaka University*

*1-3 Machikaneyama, Toyonaka, Osaka 560-8531, Japan*

*delmo@semi.ee.es.osaka-u.ac.jp

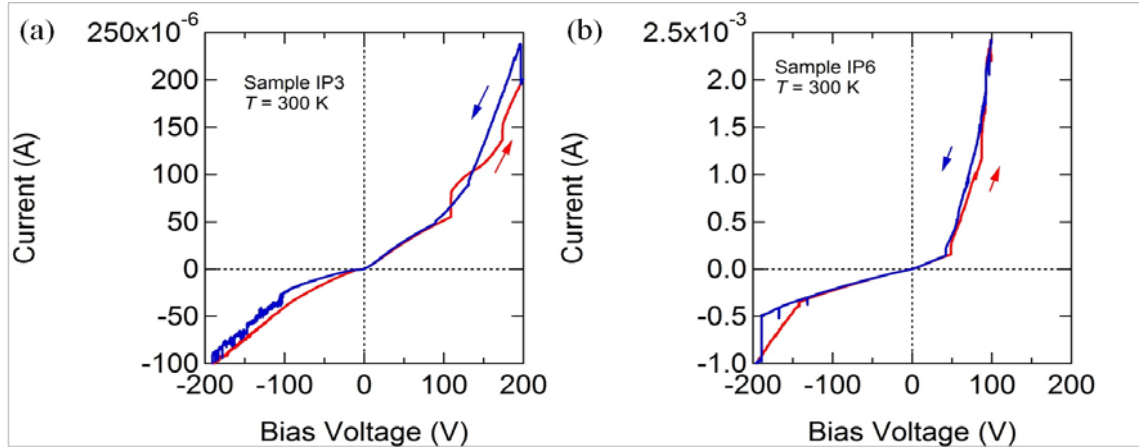

FIG. S1. Current-voltage (*I-V*) characteristics of p-Si devices at 300K and 0 T. The *I-V* curves of (a) Sample IP3 and (b) Sample IP6 are shown in forward (red line) and reverse (blue line) directions of $V_{Bias}$ from -200 to 200 V. The red and blue arrows show the direction of $V_{Bias}$. The *I-V* curves are asymmetric due to the preparation of the contact. We stress that at low $V_{Bias}$, the *I-V* is Ohmic. We think that the asymmetry of the *I-V* curves is due to the nature of the electrode/silicon contact. Although, we cautiously prepared the devices to fabricate Ohmic electrode/silicon contacts as identical as possible, this is difficult to attain because we prepared the devices manually by hand, by using diamond cutter for cleaving the silicon substrate and soldering iron to augment the indium metal to the substrates.

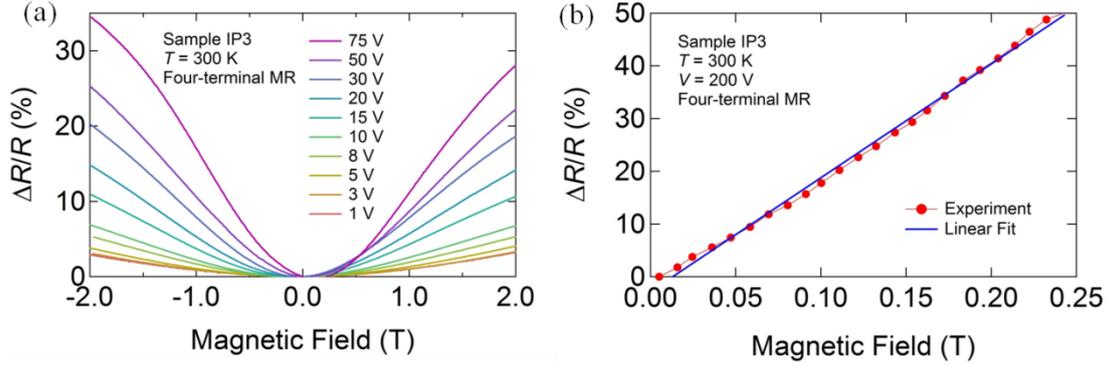

FIG. S2. Four-terminal MR of p-Si device at 300 K. (a) MR of the p-Si device, as a function of $H$ from -2 to 2 T, for a variety of $V_{Bias}$ from 1 to 75 V. (b) MR as a function of $H$ from 0 to 250 mT at 200 V (dotted red line). The linear fit (blue line) shows that the MR is linear, even in low magnetic fields.

The MR at low $H$ can be enhanced geometrically in four-terminal bipolar silicon device at room temperature, as a result of the formation of the so-called p-n boundary (see Fig. 1 and 3 in Ref. [14]), which arises from injecting electrons and holes of equal densities [14]. However, this cannot explain our results because the large MR is mainly observed at very high voltages, where electrons dominate the overall transport. Indeed, the transport in the p-Si device is bipolar at high $V_{Bias}$, because $n_e$ is comparable to $n_h$. Thus, by adopting the methods used in Ref. [14], we calculated $<\mu>$ = $(\mu_e + C\mu_h)/(1 + C) \approx 730$ cm$^2$V$^{-1}$s$^{-1}$, where $C = n_h/n_e = 0.85$, and $\Delta\mu = [C^{1/2}/(1 + C)]|\mu_e - \mu_h| \approx 350$ cm$^2$V$^{-1}$s$^{-1}$ for $V_B = 200$ V to evaluate the effect of inhomogeneity on the MR at low $H$ [14]. Since, $\Delta\mu/<\mu> = 0.48$ (less than one), this indicates that $<\mu>$ controls the MR [2,3]. However, MR $\propto <\mu>H$ $\approx 1.8$ % at 250 mT and $|H_c|$ is $<\mu>^{-1} \approx 13.5$ T, which again are very different to the experimentally obtained results in Fig. 1(b) and 1(c) of the paper. This indicates that, even in bipolar injection, the contribution of inhomogeneity to the large linear MR at low $H$ is negligibly small. Moreover, to verify experimentally the contribution of the p-n boundary to the large linear MR, we measured simultaneously the four-terminal and two-terminal MR of the device. Figure S1(a) shows the four-terminal MR as a function of $H$ from -5 to 5 T for various $V_B$ from 1 to 75 V. Figure S1(b) also shows that the four-terminal MR is linear below 250 mT at 200 V, as suggested by the linear fit (blue line). We found that the result of the four-terminal MR in Fig. S1(a) and S1(b) is consistent with that of the two-terminal MR in Fig. 1(b) and 1(c) of the paper that the MR is enhanced when electrons dominate the transport, thus suggesting negligible contribution from the p-n boundary [14]. We note that the MR in Fig. S2(b) is smaller than the two-terminal MR in Fig. 1(c) of the paper, because the separation of the voltage probes on top of the device is shorter than the length of the device. This is expected because the MR depends on the distance from the injection contact, as suggested in Fig. 3 and ref. [9].

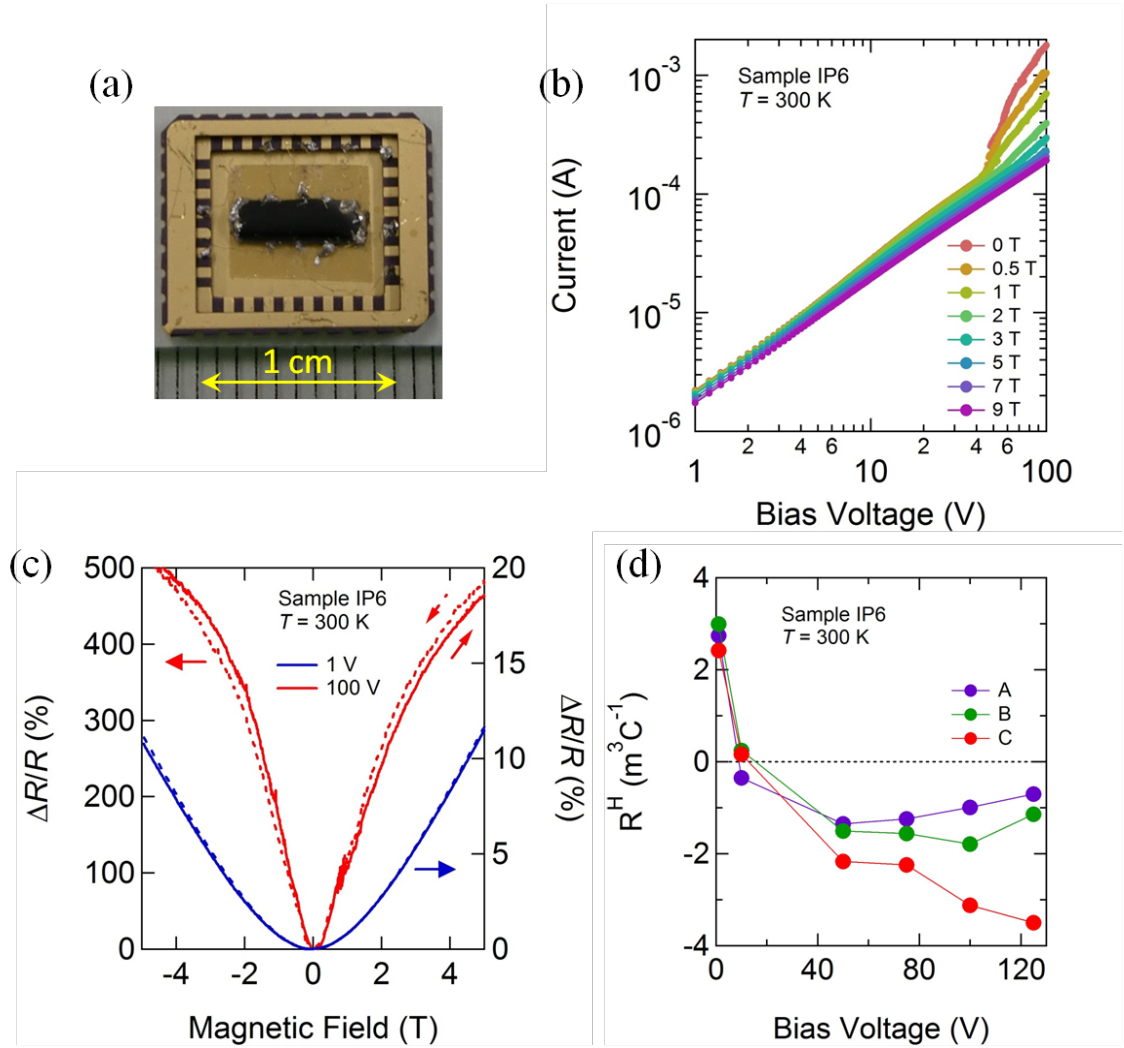

FIG. S3. Magnetotransport properties of p-Si device (Sample IP6) at 300 K. (a) Photo of the p-Si device for Hall measurement at different Hall probe positions. Three pairs of Hall electrodes were fabricated at the edges of the device for Hall voltage measurements. Fabrication method is reported in Ref. [9]. Schematic of the measurement is shown in Fig. 3(a) of the paper. (b) *I-V* characteristics of the p-Si device for various *H* from 0 to 9 T, plotted in $\log_{10}$-$\log_{10}$ scale. (c) MR as a function of *H* from -5 to 5 T for *V* = 1 V (blue line) and *V* = 100 V (red line). The solid and dashed lines show the direction of the applied magnetic field. The MR is significantly enhanced at 100 V, particularly at low *H*, compared to that at 1 V. For example, the MR at 500 mT for 100 V and 1 V are ~45 % and ~0.25 %, respectively. (d) Hall coefficient ($R^H$) as a function of bias voltage for different Hall probe positions, as shown in Fig. 3(a) of the paper. At 1 V, $R^H$ is positive indicating that holes dominate the transport, whereas at 100 V $R^H$ is negative indicating that electrons become the dominant carrier.

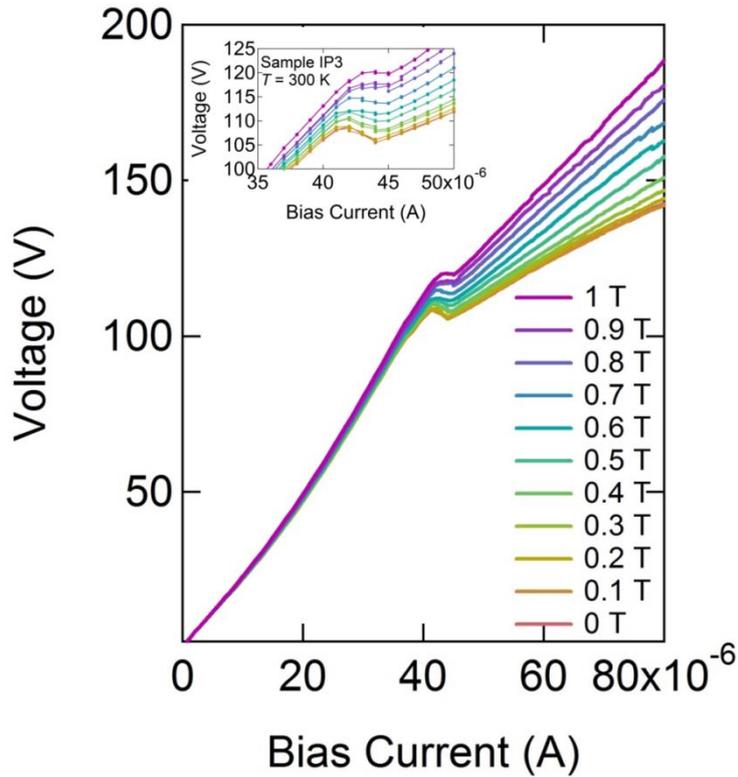

Fig. S4. We verified the existence of electron-hole plasma in our p-Si device by measuring the negative resistance, which is the decreasing of $I$ with increasing $V$ of the device [20-22,31,32]. The figure shows the current-biased $I$-$V$ of p-Si device (sample IP3) for various $H$ from 0 to 1 T. For $H =$ 0 T, the $I$-$V$ shows negative resistance in $I = 40 \sim 45$ μA regime (inset). As $H$ increases from 0 T, the $I$-$V$ shows different $H$ responses for $V$ at fixed $I$, above and below the onset ($I = 40$ μA) of the negative resistance. Above 40 μA, $V$ shows significantly large $H$ response, even for $H < 1$ T, whereas below 40 μA, $V$ is almost unchanged by $H$. Indeed, the enhancement of the MR at low $H$ arises when the plasma is formed, suggesting that the MR is controlled by both electrons and holes in the device.

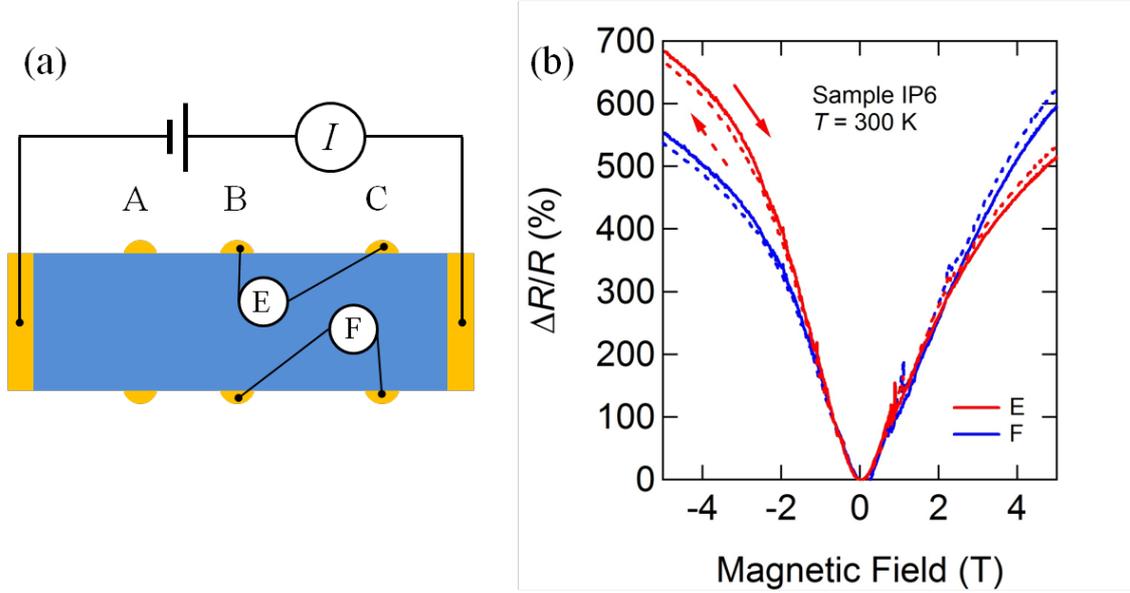

FIG. S5. Suhl effect measurement. (a) Schematic of the p-Si device (Sample IP6) and measurement setup. E and F are the voltage probes at the two edges of the device. (b) MR as a function of $H$ from -5 to 5 T measured by E (red line) and F (blue line) probes at $V_{Bias}$ = 100 V. The solid and dashed lines show the direction of the applied magnetic field. The MR is measured simultaneously in E and F voltage proves to demonstrate that the asymmetry at high $H$ is just due to the position of the probes.

As explained in ref. [32], Suhl effect can be used to demonstrate the deflection of the electron-hole plasma in magnetic field. Here, an asymmetric MR vs $H$ curve at 0 T is observed, where the magnitude of $H$ necessary for Suhl effect is comparable to the magnitude that cause the magnetic effect in the device characteristics [32]. However, in our case, the asymmetric MR was not observed, as shown in Fig. S5(b), which is comparable to the two-terminal MR in Fig. S3(c) (red line). This is expected since the electron mobility of silicon is $\mu_e \approx 3\mu_h = 1,050$ cm$^2$V$^{-1}$s$^{-1}$, which is small to induced large MR below 1 T [32]. Moreover, in the device fabrication, we see to it that the injection (source) and drain electrodes cover the whole edges of the silicon substrate, so that no isolated plasma filament is generated, which results in a significant displacement and deformation of the plasma by $H$, and hence large MR [32,37].

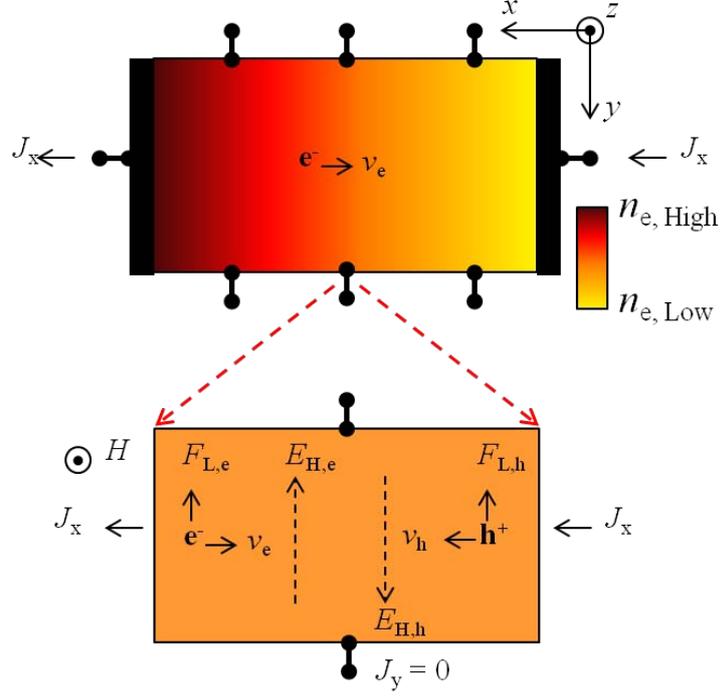

FIG. S6. Hall effect in electron-injected p-type semiconductor. Upper panel shows the schematic of the device with source-drain (left-right) electrodes, and three pairs of Hall voltage probes at the edges. The dumbbell-shaped protrusions in black represent the electrodes. Electrons (e) are injected from the left side of the device. $H$ is applied perpendicular to the plane of the paper. We assume that the electron density ($n_e$) is large around the electron injecting side and decreases as it goes toward the drain, as shown schematically by the color gradation, which is similar to that measured in Fig. 3(b) of the paper. Since, holes (h) also contribute significantly to the transport, we calculated the Hall coefficient ($R^H$) as a function of the relative mobilities ($\mu_e/\mu_h$) and relative concentration ($n_e/n_h$) of electrons and holes in an arbitrary segment of the device. Here, we assume that the device is stratified, having different resistivities for different segments [1], in which the transport in each segment is Ohmic to simplify calculations. Lower panel shows the representative segment, where electrons and holes are drifting with velocity $v_e$ and $v_h$, respectively. Here, $v_e = \mu_e E_x$ and $v_h = \mu_h E_x$, where $E_x$ is electric field in that segment.

In the presence of $H$, electrons and holes will experience Lorentz force ($F_L$) but with different magnitude because $\mu_e \neq \mu_h$, and accumulate at the same side of the device. Once equilibrium is established, there should be no net current in the y-direction ($J_y = 0$). This can be represented as

$$J_y = J_{y,e} + J_{y,h} = qn_e v_{e,y} + qn_h v_{h,y} = 0. \tag{1}$$

Here, the net force acting on electrons and holes cannot be zero because both carriers move along the *y*-direction to give $J_y = 0$. Also, the accumulation of the charges will induce Hall field ($E_H$), in which the charges will feel as Hall force ($F_H = qE_H$). Therefore, the force acting on electrons and holes are

$$F_{y,e} = F_{L,e} + F_{H,e} = -qv_e H - qE_{H,e} \ , \ F_{y,h} = F_{L,h} + F_{H,h} = -qv_h H + qE_{H,h}. \tag{2}$$

Since, $F = qv/\mu$ and $v = \mu E$, equation (2) becomes

$$\frac{v_{e,y}}{\mu_e} = E_H + \mu_e E_x H, \ \frac{v_{h,y}}{\mu_h} = -E_H + \mu_h E_x H. \tag{3}$$

From equation (1), we can substitute $-n_e v_{e,y} = n_h v_{h,y}$ to equation (3) to obtain

$$E_H(n_e \mu_e + n_h \mu_h) = E_x(n_e \mu_e^2 - n_h \mu_h^2) H. \tag{4}$$

In the *x*-direction, $J_x$ can be represented as

$$J_x = qn_e v_e + qn_e v_e = (n_e \mu_e + n_h \mu_h) qE_x. \tag{5}$$

We can use equation (5) to substitute for $E_H$ in equation (4) to obtain

$$qE_H(n_e \mu_e + n_h \mu_h)^2 = J_x(n_h \mu_h^2 - n_e \mu_e^2) H. \tag{6}$$

Since $R^H = E_y/J_x H$, equation (6) becomes

$$R^H = \frac{n_h \mu_h^2 - n_e \mu_e^2}{q(n_e \mu_e + n_h \mu_h)^2} \tag{7}$$

or

$$R^H = \frac{1 - n_r \mu_r^2}{qn_h(1 + n_r \mu_r)^2} \tag{8}$$

where $n_r = n_e/n_h$ and $\mu_r = \mu_e/\mu_h$. We used equation (8) as the fitting function in Fig. 3(c)-3(e) of the paper.